\documentclass[12pt]{extarticle}
\usepackage{setspace}

\usepackage[T1]{fontenc}

\usepackage[latin1]{inputenc}
\usepackage{epsfig}
\usepackage[english,french]{babel}
\usepackage{color}
\usepackage{graphicx}

\usepackage{dcolumn}

\usepackage{moreverb}

\usepackage{amsmath,amssymb,amsfonts}

\begin{document}
\numberwithin{equation}{section}
\newcommand{\boxedeqn}[1]{%
  \[\fbox{%
      \addtolength{\linewidth}{-2\fboxsep}%
      \addtolength{\linewidth}{-2\fboxrule}%
      \begin{minipage}{\linewidth}%
      \begin{equation}#1\end{equation}%
      \end{minipage}%
    }\]%
}

%\boxedeqn{}

\newsavebox{\fmbox}
\newenvironment{fmpage}[1]
     {\begin{lrbox}{\fmbox}\begin{minipage}{#1}}
     {\end{minipage}\end{lrbox}\fbox{\usebox{\fmbox}}}

\begin{flushleft}
\title*{{\LARGE{\textbf{Supersymmetry as a method of obtaining new superintegrable systems with higher order integrals of motion}}}}
\newline
\newline
Ian Marquette
\newline
D\'epartement de physique et Centre de recherche math\'ematique,
Universit\'e de Montr\'eal,
\newline
C.P.6128, Succursale Centre-Ville, Montr\'eal, Qu\'ebec H3C 3J7,
Canada
\newline
ian.marquette@umontreal.ca
\newline
\newline
\newline
The main result of this article is that we show that from supersymmetry we can generate new superintegrable Hamiltonians. We consider a particular case with a third order integral and apply the Mielnik's construction in supersymmetric quantum mechanics. We obtain a new superintegrable potential separable in Cartesian coordinates with a quadratic and quintic integrals and also one with a quadratic integral and an integral of order seven. We also construct a superintegrable system written in terms of the fourth Painlev\'e transcendent with a quadratic integral and an integral of order seven.
\newline
\newline
\section{Introduction}
Superintegrability [1-14] and supersymmetric quantum mechanics (SUSYQM) [15-21] have attracted a lot of attention in recent years. Both of these fields have important applications in quantum chemistry, atomic physics, molecular physics, nuclear physics and condensed matter physics. Although they are two separate issues, many quantum systems such as the harmonic oscillator, the hydrogen atom and the Smorodinsky-Winternitz potential are both superintegrable and supersymmetric [21]. Superintegrability with third order integrals was the object of a series of articles [22-26]. The systems studied have a second and a third order integrals. They were studied by means of cubic and deformed oscillator algebras. The supersymmetric quantum mechanics approach was used [25] and also higher order supersymmetric quantum mechanics [26] in order to calculate energies and wave functions. These articles indicate that superintegrability is closely connected to supersymmetry. We will show in this article that supersymmetry can provide a method of generating new superintegrable systems. We will consider two-dimensional systems separable in Cartesian coordinates. The separability implies the existence of a second order integral of motion.
\newline
\newline
Let us recall some definitions concering superintegrability and supersymmetry. In classical mechanics a Hamiltonian system with Hamiltonian H and integrals of motion $X_{a}$
\newline
\begin{equation}
H=\frac{1}{2}g_{ik}p_{i}p_{k}+V(\vec{x},\vec{p}),\quad X_{a}=f_{a}(\vec{x},\vec{p}),\quad a=1,..., n-1 \quad,
\end{equation}
\newline
is called completely integrable (or Liouville integrable) if it
allows n integrals of motion (including the Hamiltonian) that are
well defined functions on phase space, are in involution
$\{H,X_{a}\}_{p}=0$, $\{X_{a},X_{b}\}_{p}=0$, a,b=1,...,n-1 and
are functionally independent ($\{,\}_{p}$ is a Poisson bracket). A
system is superintegrable if it is integrable and allows further
integrals of motion $Y_{b}(\vec{x},\vec{p})$, $\{H,Y_{b}\}_{p}=0$,
b=n,n+1,...,n+k that are also well defined functions on phase
space and the integrals$\{H,X_{1},...,X_{n-1},Y_{n},...,Y_{n+k}\}$
are functionally independent. A system is maximally
superintegrable if the set contains 2n-1 such integrals. The integrals
$Y_{b}$ are not required to be in evolution with
$X_{1}$,...$X_{n-1}$, nor with each other. The same definitions apply in quantum mechanics but
$\{H,X_{a},Y_{b}\}$ are well defined quantum mechanical operators,
assumed to form an algebraically independent set.
\newline
\newline
\newline
In Section 2, we recall definitions and results of supersymmetric quantum mechanics. We also discuss some results obtained by B.Mielnik [27]. B.Mielnik showed that the factorization of second order operators is not necessarily unique. Supersymmetric quantum mechanics allows to find the eigenfunctions, the energy spectrum, creation and annihilation operators. In Section 3, we will consider a two-dimensional Hamiltonian consisting of two one-dimensional Hamiltonians that are superpartners. Such systems are by construction separable in Cartesian coordinates so a second order integral exists. From the creation and annihilation operators of the one dimensional part we can generate a higher order integral of motion. The system is thus superintegrable. We show how these results allow us to recover known superintegrable systems with a third order integral that are special cases of a Hamiltonian written in terms of the fourth Painlev\'e transcendent. In Section 4, we consider a particular case with a third order integral, apply the Mielnik's method and obtain a new superintegrable potential separable in Cartesian coordinates with a quadratic and quintic integrals and also one with a quadratic and seventh order integrals. We also constuct a  superintegrable system written in terms of the fourth Painlev\'e transcendent with a quadratic and seventh order integrals.
\newline
\section{Supersymmetry and Mielnik's factorization method}
We begin this Section by recalling definitions and results of supersymmetric quantum mechanics. We define two first order operators
\newline
\begin{equation}
A=\frac{\hbar}{\sqrt{2}}\frac{d}{dx}+W(x),\quad A^{\dagger}=-\frac{\hbar}{\sqrt{2}}\frac{d}{dx}+W(x) \quad .
\end{equation}
We consider the following two Hamiltonians which are called "superpartners"
\begin{equation}
H_{1}=A^{\dagger}A=-\frac{\hbar^{2}}{2}\frac{d^{2}}{dx^{2}}+W^{2}-\frac{\hbar}{\sqrt{2}}W',\quad H_{2}= AA^{\dagger}=-\frac{\hbar^{2}}{2}\frac{d^{2}}{dx^{2}}+W^{2}+\frac{\hbar}{\sqrt{2}}W'.
\end{equation}
There are two cases. The first is $A\psi_{0}^{(1)}\neq 0$, $E_{0}^{(1)}\neq 0$, $A^{\dagger}\psi_{0}^{(2)}\neq 0$ and $E_{0}^{(2)}\neq 0$. We have
\newline
\begin{equation}
E_{n}^{(2)}=E_{n}^{(1)}>0,\quad \psi_{n}^{(2)}=\frac{1}{\sqrt{E_{n}^{(1)}}}A\psi_{n}^{(1)},\quad  \psi_{n}^{(1)}=\frac{1}{\sqrt{E_{n}^{(2)}}}A^{\dagger}\psi_{n}^{(2)} \quad .
\end{equation}
and the two Hamiltonians are isospectral. This case corresponds to broken supersymmetry.
\newline
For the second case the supersymmetry is unbroken and we have $A\psi_{0}^{(1)}= 0$, $E_{0}^{(1)}= 0$, $A^{\dagger}\psi_{0}^{(2)}\neq 0$ and $E_{0}^{(2)}\neq 0$. Without lost of generality we take $H_{1}$ as having a zero energy ground state. We have
\newline
\begin{equation}
E_{n}^{(2)}=E_{n+1}^{(1)},\quad E_{0}^{(1)}=0,\quad \psi_{n}^{(2)}=\frac{1}{\sqrt{E_{n+1}^{(1)}}}A \psi_{n+1}^{(1)},\quad  \psi_{n+1}^{(1)}=\frac{1}{\sqrt{E_{n}^{(2)}}}A^{\dagger} \psi_{n}^{(2)} \quad .
\end{equation}
We can define the matrices
\begin{equation}
H = \begin{pmatrix}H_{1} & 0 \\ 0 & H_{2}\end{pmatrix}\quad Q = \begin{pmatrix}0 & 0 \\ A & 0\end{pmatrix}\quad Q^{\dagger} = \begin{pmatrix} 0 & A^{\dagger} \\ 0 & 0\end{pmatrix}\quad .
\end{equation}
\newline
They satisfy the relations 
\newline
\begin{equation}
[H,Q]=[H,Q^{\dagger}]=0, \quad  \{Q,Q\}=\{Q^{\dagger},Q^{\dagger}\}=0, \quad  \{Q,Q^{\dagger}\}=H \quad .
\end{equation}
\newline
The operators $Q$, $Q^{\dagger}$ are called "supercharges". We have a sl(1|1) superalgebra and $H_{1}$ and $H_{2}$ are superpartners. Supersymmetric quantum mechanics allow us to obtain  the creation and annihilation operators. The operators $M^{\dagger}$ and $M$ with $b^{\dagger}$ and $b$ respectively the creation and annihilation operators for the Hamiltonian $H_{1}$
\newline
\begin{equation}
M=A^{\dagger}bA,\quad M^{\dagger}=A^{\dagger}b^{\dagger}A,
\end{equation}
are thus the creation and annihilation operators for the Hamiltonian $H_{2}$.
\newline
Supersymmetric quantum mechanics with higher order supercharges has been studied [28-32]. The case with second order operators of the form
\begin{equation}
M^{\dagger}=\partial^{2}-2h(x)\partial+c(x),\quad M=\partial^{2}+2h(x)\partial+c(x)\quad .
\end{equation}
was investigated. The case with a first and second order supersymmetry was also treated.
\newline
\newline
As far as we could find, the generalized ladder operators appeared first in Deift [33], but we shall follow a somewhat different approach. We present the further results in supersymmetric quantum mechanics by recalling results obtained by B.Mielnik [27] concerning the search of superpartners for the harmonic oscillator. He pointed out that the factorization is not unique. He presented a new derivation of a important class of potentials previously obtained by P.B.Abraham and H.E.Moses with the Gelfand-Levitan formalism [34]. Their energy and eigenfunctions can be directly obtained from the harmonic oscillator up to a zero mode state. In Section 3, we will show how this family of Hamiltonians is related to superintegrable systems with third order integrals.
\newline
\newline
We consider the following Hamiltonian
\newline
\begin{equation}
H_{osc}=\frac{1}{2}\frac{d^{2}}{dx^{2}}+\frac{x^{2}}{2}\quad .
\end{equation}
We introduce the following first order operators
\begin{equation}
a=\frac{1}{\sqrt{2}}(\frac{d}{dx}+x),\quad a^{\dagger}=\frac{1}{\sqrt{2}}(-\frac{d}{dx}+x) \quad .
\end{equation}
The Hamiltonian $H_{1}$ and $H_{2}$ are superpartners and have in fact the shape invariance properties
\newline
\begin{equation}
a^{\dagger}a=H_{osc}-\frac{1}{2}=H_{1}, \quad aa^{\dagger}=H_{osc}+\frac{1}{2}=H_{2} \quad .
\end{equation}
\newline
This construction allows us to find the energy spectrum and the eigenfunction algebraically. B.Mielnik considered the Hamiltonian $H_{2}$ and showed that the operators $a$ and $a^{\dagger}$ are not unique [27]. He defined the following new operators
\newline
\begin{equation}
b=\frac{1}{\sqrt{2}}(\frac{d}{dx}+\beta(x)), \quad b^{\dagger}=\frac{1}{\sqrt{2}}(-\frac{d}{dx}+\beta(x)),
\end{equation}
and required
\newline
\begin{equation}
H_{2}=H_{osc}+\frac{1}{2}=bb^{\dagger}\quad .
\end{equation}
He obtained the following Riccati equation [35]
\newline
\begin{equation}
\beta'(x)+\beta^{2}(x)=1+x^{2}\quad .
\end{equation}
The fact of knowing a particular solution $(\beta(x)=x)$ allows to find the general solution [35]. He defined
\newline
\begin{equation}
\beta(x)=x+\phi(x)\quad ,
\end{equation}
and found
\newline
\begin{equation}
\phi(x)=\frac{e^{-x^{2}}}{\gamma+\int_{0}^{x}e^{-x'^{2}}dx'}\quad .
\end{equation}
where $\gamma$ is a constant. There are two cases: with a singularity and without singularity. 
The inverted product $b^{\dagger}b$ was not $H_{2}$+const and was a new Hamiltonian
\newline
\begin{equation}
H'=b^{\dagger}b=H_{2}-\phi'(x)=-\frac{1}{2}\frac{d^{2}}{dx^{2}}+\frac{x^{2}}{2}-\frac{d}{dx}( \frac{e^{-x^{2}}}{\gamma+\int_{0}^{x}e^{-x'^{2}}dx'} )\quad .
\end{equation}
We can obtain from $H_{2}$ the creation and annihilation operators for $H'$. These operators are given by the following expression
\newline
\begin{equation} 
s^{\dagger}=b^{\dagger}a^{\dagger}b,\quad s=b^{\dagger}ab \quad ,
\end{equation}
with $a$ and $a^{\dagger}$ the annihilation and creation operators for $H_{2}$. The eigenfunctions and energy spectrum of the Hamiltonian H' can be obtained from the Eq.(2.4). The coherent states have also been studied extensively [36]. This system is a special case of a one dimensional part of a Hamiltonian separable in Cartesian coordinates written in terms of the fourth Painlev\'e transcendent. 
\newline
\section{Higher order integrals of motion and SUSYQM}
Let us consider a two-dimensional Hamiltonian separable in Cartesian coordinates $H_{t}(x,y,P_{x},P_{y})=H_{x}(x,P_{x})+H_{y}(y,P_{y})$ with creation and annihilation operators (polynomial in momenta) $A_{x}$, $A_{x}^{\dagger}$, $A_{y}$ and $A_{y}^{\dagger}$. These operators satisfy
\newline
\begin{equation}
[H_{x},A_{x}^{\dagger}]=\lambda_{x}A_{x}^{\dagger},\quad [H_{y},A_{y}^{\dagger}]=\lambda_{y}A_{y}^{\dagger}\quad .
\end{equation}
The following operators
\newline
\begin{equation}
f_{1}=A_{x}^{\dagger m}A_{y}^{n},\quad f_{2}=A_{x}^{m}A_{y}^{\dagger n}\quad ,
\end{equation}
commute with the Hamiltonian H
\newline
\begin{equation}
[H_{t},f_{1}]=[H_{t},f_{2}]=0
\end{equation}
if 
\newline
\begin{equation}
m\lambda_{x}-n\lambda_{y}=0, \quad  m,n \in \mathbb{Z}^{+} \quad .
\end{equation}
\newline
Creation and annihilation operators allow us to construct polynomial integrals of motion.
\newline
The following sums are also polynomial integrals that commute with the Hamiltonian H
\newline
\begin{equation}
I_{1}=A_{x}^{\dagger m}A_{y}^{n}- A_{x}^{m}A_{y}^{\dagger n}, \quad I_{2}=A_{x}^{\dagger m}A_{y}^{n}+ A_{x}^{m}A_{y}^{\dagger n}\quad .
\end{equation}
\newline
There are the integrals $I_{1}$ and $I_{2}$. The system $H_{t}$ is thus superintegrable. By construction, the Hamiltonian $H_{t}$ possesses a second order integral ($K=H_{x}-H_{y}$). The integral $I_{2}$ is the commutator of $I_{1}$ and $K$. The Hamiltonian $H_{t}$ is thus superintegrable. We will show how supersymmetry makes it possible to construct superintegrable systems from one-dimensional Hamiltonian $H_{x}$ with creation and annihilation operators $A_{x}^{\dagger}$ and $A_{x}$. We choose in the y-axis a superpartner (or a family of superpartners). This Hamiltonian $H_{y}$ possess creation and annihilation operators that can be obtain from Eq.(2.7). A direct consequence of supersymmetry is the relation $\lambda_{x}=\lambda_{y}$. We have thus the following integrals
\newline
\begin{equation}
K=H_{x}-H_{y},\quad I_{1}=A_{x}^{\dagger}A_{y}- A_{x}A_{y}^{\dagger},\quad I_{2}=A_{x}^{\dagger}A_{y}+ A_{x}A_{y}^{\dagger}\quad .
\end{equation}
\newline
Let us, apply this construction to the interesting systems found by Mielnik. We take in the x axis the Hamiltonian $H_{2}$ given by Eq.(2.9) and in the y axis its superpartner $H'$ given by Eq.(2.17). We obtain a superintegrable system with integrals  given by Eq.(3.6) with Eq.(2.10) and (2.18) 
\newline
\begin{equation}
K=H_{x}-H_{y},\quad I_{1}=a_{x}^{\dagger}s_{y}-a_{x}s_{y}^{\dagger},\quad I_{2}=a_{x}s_{y}^{\dagger}+a_{x}^{\dagger}s_{y}\quad ,
\end{equation} 
where $a^{\dagger}_{x}$, $a_{x}$, $s_{y}^{\dagger}$ and $s_{y}$ are respectively the creation and annihilation operators of $H_{2}$ and $H'$.
\newline
These integrals are of order 2, 3 and 4. This superintegrable system appears in the investigation of superintegrable systems with a second and a third order integrals separable in Cartesian coordinates. This is a particular case of a Hamiltonian written in terms of the fourth Painlev\'e transcendent found by S.Gravel [23] and studied in Ref. 26. 
\newline
\section{Construction of new superintegrable systems}
\subsection{Hamiltonians involving the error function}
We consider the following superintegrable systems obtained in Ref. 23 and studied in Ref. 25 from the point of view of cubic algebras and SUSYQM
\newline
\begin{equation}
H_{g}=-\frac{\hbar^{2}}{2}(\frac{d^{2}}{dx^{2}}+\frac{d^{2}}{dy^{2}}) +\hbar^{2}[\frac{x^{2}+y^{2}}{8a^{4}} + \frac{1}{(x-a)^{2}}+\frac{1}{(x+a)^{2}}].
\end{equation}
We consider the case $a=ia_{0}, a_{0}\in \mathbb{R}$. Let us define the two operators 
\newline
\newline
\begin{equation}
c^{\dagger}= \frac{1}{\sqrt{2}}(-\hbar \frac{d}{dx} + \frac{ \hbar}{2a_{0}^{2}}x
+\hbar (\frac{1}{x-ia_{0}}+\frac{1}{x+ia_{0}})),
\end{equation}
\begin{equation}
c= \frac{1}{\sqrt{2}}(\hbar \frac{d}{dx} + \frac{
\hbar}{2a_{0}^{2}}x +\hbar (\frac{1}{x-ia_{0}}+\frac{1}{x+ia_{0}}))\quad .
\end{equation}
We have 
\newline
\begin{equation}
H_{s1}=b^{\dagger}b=\frac{P_{x}^{2}}{2}+\frac{\hbar^{2}x^{2}}{8a_{0}^{4}}+\frac{\hbar^{2}}{(x-ia_{0})^{2}}+\frac{\hbar^{2}}{(x+ia_{0})^{2}}+\frac{3\hbar^{2}}{4a_{0}^{2}},
\end{equation}
\begin{equation}
H_{s2}=bb^{\dagger}=\frac{P_{x}^{2}}{2}+\frac{\hbar^{2} x^{2}}{8 a_{0}^{4}}+
\frac{5\hbar^{2}}{4a_{0}^{2}}\quad .
\end{equation}
The Hamiltonian $H_{g}$ is the sum up to a constant of $H_{s1}$ and $H_{s2}$. We apply the Mielnik's procedure to the Hamiltonian $H_{s1}$ to find all the superpartners. We define the following operator
\newline
\begin{equation}
d=\frac{\hbar}{2}(\frac{d}{dx}+\beta(x)), \quad d^{\dagger}=\frac{\hbar}{2}(-\frac{d}{dx}+\beta(x)), 
\end{equation}
and demand $H_{s1}=d^{\dagger}d$. We obtain the following Riccati equation
\newline
\begin{equation}
\beta'(x)+\beta^{2}(x)=\frac{\hbar^{2}x^{2}}{8a_{0}^{4}}+ \frac{\hbar^{2}}{(x-ia_{0})^{2}}+ \frac{\hbar^{2}}{(x+ia_{0})^{2}} +\frac{3\hbar^{2}}{4a_{0}^{2}}\quad .
\end{equation}
We know a particular solution
\newline
\begin{equation}
\beta_{0}= \frac{1}{2a_{0}^{2}}x+ (\frac{1}{x-ia_{0}}+\frac{1}{x+ia_{0}})\quad .
\end{equation}
Because we know a particular solution we can found the general solution. We consider
\begin{equation}
\beta=\beta_{0}(x)+\phi(x)\quad ,
\end{equation}
and obtain the following the following equation 
\begin{equation}
\phi'(x)+\phi^{2}(x)+2\beta_{0}(x)\phi(x)=0 \quad .
\end{equation}
We consider the transformation $z(x)=\frac{1}{\phi(x)}$ and obtain a first order linear inhomogeneous equation
\newline
\begin{equation}
-z'(x)+2\beta_{0}(x)z(x)+1=0 \quad .
\end{equation}
We obtain
\begin{equation}
z(x)=e^{\frac{x^{2}}{2a_{0}^{2}}}(a_{0}^{2}+x^{2})^{2}\gamma+\frac{1}{4a_{0}^{3}}(a_{0}^{2}+x^{2})(2a_{0}x+e^{\frac{x^{2}}{2a_{0}^{2}}}\sqrt{2\pi}(a_{0}^{2}+x^{2})Erf(\frac{x}{\sqrt{2}a_{0}}))
\end{equation}
\begin{equation}
\beta(x)= \frac{1}{2a_{0}^{2}}x+(\frac{1}{x-ia_{0}}+\frac{1}{x+ia_{0}})+
\end{equation}
\[ \frac{1}{e^{\frac{x^{2}}{2a_{0}^{2}}}(a_{0}^{2}+x^{2})^{2}\gamma+\frac{1}{4a_{0}^{3}}(a_{0}^{2}+ x^{2})(2a_{0}x+e^{\frac{x^{2}}{2a_{0}^{2}}}\sqrt{2\pi}(a_{0}^{2}+x^{2})Erf(\frac{x}{\sqrt{2}a_{0}}))}   .          \]
Using the function z(x) given by the Eq.(4.12) the family of superpartner is thus given by
\newline
\begin{equation}
H_{\gamma}=H_{s1}-\phi'(x)=\frac{P_{x}^{2}}{2}+\frac{\hbar^{2}x^{2}}{8a_{0}^{4}}+\frac{\hbar^{2}}{(x-ia_{0})^{2}}+\frac{\hbar^{2}}{(x+ia_{0})^{2}}+\frac{3\hbar^{2}}{4a_{0}^{2}}
\end{equation}
\[-\frac{d}{dx}[ \frac{1}{e^{\frac{x^{2}}{2a_{0}^{2}}}(a_{0}^{2}+x^{2})^{2}\gamma+\frac{1}{4a_{0}^{3}}(a_{0}^{2}+x^{2})(2a_{0}x+e^{\frac{x^{2}}{2a_{0}^{2}}}\sqrt{2\pi}(a_{0}^{2}+x^{2})Erf(\frac{x}{\sqrt{2}a_{0}}))}  ].\]
\newline
The eigenfunctions and energy spectrum of Hamiltonian $H_{s_{1}}$ have been obtained in Ref. 25 from supersymmetry.
The eigenfunctions and energy spectrum of $H_{\gamma}$ can be obtained directly from $H_{s_{1}}$ and Eq.(2.4). We can also obtain the creation and annihilation operators from those of $H_{s1}$. If we take $H_{x}=H_{s1}$ and $H_{y}=H_{\gamma}$ (the Hamiltonian $H_{\gamma}$ is thus now given in term of the variable y) we obtain a new superintegrable Hamiltonian:
\newline
\begin{equation}
H_{e}=H_{x}+H_{y}=\frac{P_{x}^{2}}{2}+ \frac{P_{y}^{2}}{2}+\frac{\hbar^{2}y^{2}}{8a_{0}^{4}}+\frac{\hbar^{2}}{(y-ia_{0})^{2}}+ \frac{\hbar^{2}}{(y+ia_{0})^{2}}+\frac{3\hbar^{2}}{4a_{0}^{2}}
\end{equation}
\[-\frac{d}{dy}[ \frac{1}{e^{\frac{y^{2}}{2a_{0}^{2}}}(a_{0}^{2}+y^{2})^{2}\gamma+\frac{1}{4a_{0}^{3}}(a_{0}^{2}+y^{2})(2a_{0}y+e^{\frac{y^{2}}{2a_{0}^{2}}}\sqrt{2\pi}(a_{0}^{2}+y^{2})Erf(\frac{y}{\sqrt{2}a_{0}}))}  ]\]
\[ +\frac{\hbar^{2}x^{2}}{8a_{0}^{4}}+\frac{\hbar^{2}}{(x-ia_{0})^{2}}+ \frac{\hbar^{2}}{(x+ia_{0})^{2}}+\frac{3\hbar^{2}}{4a_{0}^{2}}\quad . \]
\newline
The creation and annihilation operators for the Hamiltonian $H_{s1}$ are
\newline
\begin{equation}
m_{x}^{\dagger}=c_{x}^{\dagger}a_{x}^{\dagger}c_{x},\quad m_{x}=c_{x}^{\dagger}a_{x}c_{x},
\end{equation}
with
\begin{equation}
a_{x}= \frac{\hbar}{2a_{0}^{2}}(x+2a_{0}^{2}\frac{d}{dx}),\quad a_{x}^{\dagger}= \frac{\hbar}{2a_{0}^{2}}(x-2a_{0}^{2}\frac{d}{dx})\quad .
\end{equation}
\newline
The creation and annihilation operators of the Hamiltonian $H_{\gamma}$ are
\newline
\begin{equation}
r_{y}^{\dagger}=d_{y}^{\dagger}m_{y}^{\dagger}d_{y},\quad r_{y}=d_{y}^{\dagger}m_{y}d_{y}\quad .
\end{equation}
We can find from Eq.(3.6) the integrals of motion of the Hamiltonian $H_{e}$ of the order 2, 7 and 8
\newline
\begin{equation}
K=H_{x}-H_{y},\quad I_{1}=m_{x}^{\dagger}r_{y}-m_{x}r_{y}^{\dagger},\quad I_{2}=m_{x}^{\dagger}r_{y}+m_{x}r_{y}^{\dagger}\quad .
\end{equation}
The integral $I_{2}$ is given by the commutator of the integrals K and $I_{1}$.
\newline
Because the harmonic oscillator is also isospectral to $H_{\gamma}$ we have also the following superintegrable systems where we take $H_{x}=H_{s2}$ and $H_{y}=H_{\gamma}$
\newline
\begin{equation}
H_{f}=H_{x}+H_{y}=\frac{P_{x}^{2}}{2}+ \frac{P_{y}^{2}}{2}+\frac{\hbar^{2}x^{2}}{8a_{0}^{4}}+\frac{\hbar^{2}y^{2}}{8a_{0}^{4}}+\frac{\hbar^{2}}{(y-ia_{0})^{2}}+ \frac{\hbar^{2}}{(y+ia_{0})^{2}}+\frac{9\hbar^{2}}{4a_{0}^{2}}
\end{equation}
\[-\frac{d}{dy}[ \frac{1}{e^{\frac{y^{2}}{2a_{0}^{2}}}(a_{0}^{2}+y^{2})^{2}\gamma+\frac{1}{4a_{0}^{3}}(a_{0}^{2}+y^{2})(2a_{0}y+e^{\frac{y^{2}}{2a_{0}^{2}}}\sqrt{2\pi}(a_{0}^{2}+y^{2})Erf(\frac{y}{\sqrt{2}a_{0}}))}  ].\]
\newline
We have from Eq.(3.6) the following integrals of order 2, 5 and 6
\begin{equation}
K=H_{x}-H_{y},\quad I_{1}=a_{x}^{\dagger}r_{y}-a_{x}r_{y}^{\dagger},\quad I_{2}=a_{x}^{\dagger}r_{y}+a_{x}r_{y}^{\dagger}.
\end{equation}
The integral $I_{2}$ is given by the commutator of the integrals K and $I_{1}$.
\newline
\subsection{Hamiltonians with fourth Painlev\'e transcendent}
The following superintegrable system written in terms of the fourth Painlev\'e transcendent can also be related to supersymmetric quantum mechanics [26]
\newline
\begin{equation}
H_{p1}=\frac{P_{x}^{2}}{2}+\frac{P_{y}^{2}}{2}+g_{1}(x)+g_{2}(y) \quad ,
\end{equation}
\begin{equation}
g_{1}(x)=\frac{\omega^{2}}{2}x^{2}+\epsilon\frac{\hbar\omega}{2}f^{'}(\sqrt{\frac{\omega}{\hbar}}x)+\frac{\omega\hbar}{2}f^{2}(\sqrt{\frac{\omega}{\hbar}}x)+\omega \sqrt{\hbar \omega}xf(\sqrt{\frac{\omega}{\hbar}}x)+\frac{\hbar\omega}{3}(-\alpha+\epsilon) \quad ,            
\end{equation}
\begin{equation}
g_{2}(y)=\frac{\omega^{2}}{2}y^{2} \quad .
\end{equation}
\newline
This Hamiltonian has a second and third order integrals.
\newline
The function f is the fourth Painlev\'e transcendent and $f'=\frac{df}{dz}$, $z=\sqrt{\frac{\omega}{\hbar}}x$
\begin{equation}
f^{''}(z) = \frac{f^{'2}(z)}{2f(z)} + \frac{3}{2}f^{3}(z) + 4zf^{2}(z) + 2(z^{2} -
\alpha)f(z) +  \frac{\beta}{f(z)} \quad ,
\end{equation}
\newline
\begin{equation}
f(z)=P_{4}(z,\alpha,\beta).
\end{equation}
\newline
\newline
We will show that we can find new superintegrable systems from the superpartners of a one dimensional Hamiltonian with potential $g_{1}$ given by Eq.(4.23). This system was discussed in Ref. 26 and 30 and has a first and second order supersymmetry that allow to get the eigenfunctions and the energy spectrum. This system can have three, two or one infinite sequence of levels depending on parameters $\alpha$ and $\beta$. When a potential possesses only one infinite sequence of energies, this potential may also allow singlet or doublet states.
\newline
\newline
Let us consider
\begin{equation}
H_{i}=P_{x}^{2}+V_{i}(x)\quad ,\quad i=1,2.
\end{equation}
with a supersymmetry of order 1 and 2 with the following operators
\begin{equation}
q^{\dagger}=\frac{\hbar}{\sqrt{2}}(\partial + W(x)),\quad q=-\frac{\hbar}{\sqrt{2}}(\partial +W(x))\quad ,
\end{equation}
\begin{equation}
M^{\dagger}=\partial^{2}-2h(x)\partial+b(x),\quad M=\partial^{2}+2h(x)\partial+b(x)\quad .
\end{equation}
From first order supersymmetry we have
\begin{equation}
V_{1}=W'(x)+W^{2}(x),\quad V_{2}=-W'(x)+W^{2}(x)-\frac{2\omega}{\hbar},
\end{equation}
(another relations can be obtained from the supersymmetry of second order). The compatibility condition leads to 
\newline
\begin{equation}
W(x)=-h(x)-\sqrt{\frac{\omega}{\hbar}}x  \quad .
\end{equation}
\newline
The potentials $V_{1}$ and $V_{2}$ are obtained from (4.23) putting respectively $\epsilon=-1$ and $\epsilon=1$ and adding $\hbar\omega(\frac{\alpha}{3}-\frac{\epsilon}{3}-1)$ ( with $h(x)=\sqrt{\frac{\omega}{\hbar}}f(\sqrt{\frac{\omega}{\hbar}}x)$). We can apply the method to the Hamiltonian $H_{1}(x)$ and find new operators $k^{\dagger}$ and $k$ that factorize $H_{1}$.
\newline
\begin{equation}
k=\frac{\hbar}{2}(\frac{d}{dx}+\beta(x)), \quad k^{\dagger}=\frac{\hbar}{2}(-\frac{d}{dx}+\beta(x)), 
\end{equation}
This leads to a Riccati equation that we can solve because we know the particular solution $W(x)$ and we find
\newline
\begin{equation}
z(x)=\frac{1}{\phi(x)}=e^{\int^{x} 2W(x')dx'}(\gamma + \int^{x}e^{\int^{x'}2W(x'')dx''}dx'),
\end{equation}
\begin{equation}
\beta(x)=W(x)+\frac{1}{z(x)}\quad .
\end{equation}
We obtain
\newline
\begin{equation}
H_{susy}=\frac{P_{x}^{2}}{2}-\frac{d}{dx}(\phi(x))
\end{equation}
\[+ \frac{\omega^{2}}{2}x^{2}- \frac{\hbar\omega}{2}f^{'}(\sqrt{\frac{\omega}{\hbar}}x)+ \frac{\omega\hbar}{2}f^{2}(\sqrt{\frac{\omega}{\hbar}}x)+ \omega \sqrt{\hbar \omega}xf(\sqrt{\frac{\omega}{\hbar}}x) -\hbar\omega.\]
\newline
The eigenfunctions and the corresponding energy spectrum of $H_{1}$, $H_{2}$ and thus $H_{p_{1}}$ were discussed in Ref. 26 and 30. Thus we can obtain directly with Eq.(2.4) eigenfunctions and corresponding energy spectrum of Hamiltonian $H_{susy}$ given by Eq.(4.35). We also know the creation and annihilation operators of the Hamiltonian $H_{1}$ (and $H_{2}$) and we can obtain from them the creation and annihilation operators for $H_{susy}$ by the supersymmetry. From these operators, we can form two integrals of motion and we have from the separation of variables in Cartesian coordinates an integral of order 2. This system is superintegrable.
\newline
\newline
The creation and annihilation operators of $H_{1}$ are given by the following third order operators
\newline
\begin{equation}
a^{\dagger}=q^{\dagger}M^{\dagger},\quad a=M^{\dagger}q\quad ,
\end{equation}
\begin{equation}
M^{\dagger}=(\frac{d}{dx}+W_{1})(\frac{d}{dx}+W_{2}),\quad M=(-\frac{d}{dx}+W_{1})(-\frac{d}{dx}+W_{2}),
\end{equation}
with
\begin{equation}
W_{1,2}=-\frac{1}{2}\sqrt{\frac{\omega}{\hbar}}f(\sqrt{\frac{\omega}{\hbar}}x)\pm (\frac{\frac{1}{2}\sqrt{\frac{\omega}{\hbar}}f'(\sqrt{\frac{\omega}{\hbar}}x)-\sqrt{-\beta}\frac{\omega}{\sqrt{2}\hbar}}{\frac{1}{2}\sqrt{\frac{\omega}{\hbar}}f(\sqrt{\frac{\omega}{\hbar}}x)}   )\quad .
\end{equation}
\newline
The creation and annihilation operators of $H_{susy}$ are given by
\newline
\begin{equation}
v^{\dagger}=k^{\dagger}a^{\dagger}k,\quad v=k^{\dagger}ak\quad .
\end{equation}
\newline
The operators $v^{\dagger}$ and $v$ are quintic operators. If we take $H_{x}(x)=H_{1}$ and $H_{y}(y)=H_{susy}$ we obtain the following Hamiltonian 
\newline
\begin{equation}
H_{ss}=\frac{P_{x}^{2}}{2}+\frac{\omega^{2}}{2}x^{2}- \frac{\hbar\omega}{2}f^{'}(\sqrt{\frac{\omega}{\hbar}}x)+ \frac{\omega\hbar}{2}f^{2}(\sqrt{\frac{\omega}{\hbar}}x)+ \omega \sqrt{\hbar \omega}xf(\sqrt{\frac{\omega}{\hbar}}x) -\hbar\omega
\end{equation}
\[-\frac{d}{dy}(\phi(y))+\frac{\omega^{2}}{2}y^{2}- \frac{\hbar\omega}{2}f^{'}(\sqrt{\frac{\omega}{\hbar}}y)+ \frac{\omega\hbar}{2}f^{2}(\sqrt{\frac{\omega}{\hbar}}y)+ \omega \sqrt{\hbar \omega}yf(\sqrt{\frac{\omega}{\hbar}}y) -\hbar\omega,\]
\newline
with the integrals of motion
\begin{equation}
K=H_{x}-H_{y},\quad I_{1}=a_{x}^{\dagger}v_{y}-a_{x}v_{y}^{\dagger},\quad I_{2}=a_{x}^{\dagger}v_{y}+a_{x}v_{y}^{\dagger}\quad .
\end{equation}
The integral $I_{2}$ is given by the commutator of the integrals K and $I_{1}$. The integral $K$ is of order 2, $I_{1}$ is of order 7 and $I_{2}$ of order 8.
\newline
\section{Conclusion}
In this article, we showed how supersymmetric quantum mechanics gives us a method of obtaining new superintegrable systems with higher order integrals of motion. Supersymmetry in quantum mechanics make it possible to find eigenfunctions and energy spectra from a superpartner using the Eq. (2.3) and (2.4). From a one-dimensional Hamiltonian and its superpartner we have constructed a two-dimensional superintegrable system and its integrals. The integrals are given by the Eq.(3.6).
\newline
We discussed results obtained by B.Mielnik [27] in context of SUSYQM. We showed how we can generate a superintegrable system from the Hamiltonian he obtained and recover a particular case of a system with a third order integral from the Ref. 23 and studied in Ref. 26.
\newline
From the method, we have explicitely constructed superintegrable systems written in terms of the error function and the fourth Painlev\'e transcendent. These systems have higher integrals of motion. They possess respectively a second and a quintic integrals and a second and seventh order one. The supersymmetry allows also to find the wave functions and the energy spectrum.
\newline
This method of generating superintegrable systems can be applied to other systems obtained in the contex of supersymmetric quantum mechanics. The results can be generalized in higher dimensions.
\newline
\newline
\textbf{Acknowledgments} The research of I.M. was supported by a postdoctoral
research fellowship from FQRNT of Quebec. The author thanks P.Winternitz for very helpful comments and discussions.

\section{\textbf{References}}

1. V.Fock, Z.Phys. 98, 145-154 (1935).
\newline
2. V.Bargmann, Z.Phys. 99, 576-582 (1936).
\newline
3. J.M.Jauch and E.L.Hill, Phys.Rev. 57, 641-645 (1940).
\newline
4. M.Moshinsky and Yu.F.Smirnov, The Harmonic Oscillator In Modern
Physics, (Harwood, Amsterdam, 1966).
\newline
5. J.Fris, V.Mandrosov, Ya.A.Smorodinsky, M.Uhlir and P.Winternitz, Phys.Lett. 16, 354-356 (1965).
\newline
6. P.Winternitz, Ya.A.Smorodinsky, M.Uhlir and I.Fris, Yad.Fiz. 4,
625-635 (1966). (English translation in Sov. J.Nucl.Phys. 4,
444-450 (1967)).
\newline
7. A.Makarov, Kh. Valiev, Ya.A.Smorodinsky and P.Winternitz, Nuovo Cim. A52, 1061-1084 (1967).
\newline
8. N.W.Evans, Phys.Rev. A41, 5666-5676 (1990), J.Math.Phys. 32,
3369-3375 (1991).
\newline
9. E.G.Kalnins, J.M.Kress, W.Miller Jr and P.Winternitz,
J.Math.Phys. 44(12) 5811-5848 (2003).
\newline
10. E.G.Kalnins, W.Miller Jr and G.S.Pogosyan, J.Math.Phys. A34,
4705-4720 (2001).
\newline
11. E.G.Kalnins, J.M.Kress and W.Miller Jr, J.Math.Phys. 46,
053509 (2005), 46, 053510 (2005), 46, 103507 (2005), 47, 043514
(2006), 47, 043514 (2006).
\newline
12. E.G.Kalnins, W.Miller Jr and G.S.Pogosyan, J.Math.Phys. 47,
033502.1-30 (2006), 48, 023503.1-20 (2007).
\newline
13. P.Winternitz and I.Yurdusen, J. Math. Phys. 47, 103509 (2006).
\newline
14. J.Berube and P.Winternitz, J.Math.Phys. 45(5), 1959-1973 (2004).
\newline
15. G.Darboux, C.R.Acad.Sci. Paris, 94, 1459 (1882)
\newline
16. T.F.Moutard, C.R.Acad.Sci. Paris, 80, 729 (1875), J.de L'\'ecole Politech., 45, 1 (1879).
\newline
17. E.Schrodinger, Proc.Roy. Irish Acad., 46A, 9 (1940), 47A, 53 (1941).
\newline
18. L.Infeld and T.E.Hull, Rev.Mod.Phys., 23, 21 (1951).
\newline
19. E.Witten, Nucl.Phys. B188, 513 (1981); E.Witten, Nucl.Phys. B202, 253-316 (1982).
\newline
20. L.Gendenshtein, JETP Lett., 38, 356 (1983).
\newline
21. G.Junker, Supersymmetric Methods in Quantum and Statistical Physics, Springer, New York, (1995).
\newline
22. S.Gravel and P.Winternitz, J.Math.Phys. 43(12), 5902 (2002).
\newline
23. S.Gravel, J.Math.Phys. 45(3), 1003-1019 (2004).
\newline
24. I.Marquette and P.Winternitz, J. Phys. A: Math. Theor. 41, 304031 (2008).
\newline
25. I.Marquette, J. Math. Phys. 50, 012101 (2009).
\newline
26. I.Marquette, J.Math.Phys. 50 095202 (2009).
\newline
27. B. Mielnik, J.Math.Phys. 25 (12) 3387 1984.
\newline
28. A.Andrianov, M.Ioffe and V.P.Spiridonov, Phys.Lett. A174, 273 (1993). 
\newline
29. A.Andrianov, F.Cannata, J.P.Dedonder and M.Ioffe, Int.Mod.Phys.A10, 2683-2702 (1995).
\newline
30. A.Andrianov, F.Cannata, M.Ioffe and D.Nishnianidze, Phys.Lett.A, 266,341-349 (2000).
\newline
31. M.Plyushchay, Int.J.Mod.Phys. A15, 3679 (2000)
\newline
32. D.J.Fern\'andez et V.Hussin, J.Phys. A 32  3603-3619 (1999)
\newline
33. P.A. Deift, Duke Math. J. 45 267-310 (1978). 
\newline
34. P.B.Abraham and H.E.Moses, Phys. Rev. A 22, 1333 (1980).
\newline
35. E.L.Ince, Ordinary Differential Equations (Dover, New York,
1944).
\newline
36. D.J.Fern\'andez, V.Hussin et L.M.Nieto, J.Phys. A 27 3547-3564 (1994).

\end{flushleft}
\end{document}